\documentclass[sn-standardnature]{sn-jnl}% Standard Nature Portfolio Reference Style
\usepackage{ulem}
\usepackage{subfigure}

\jyear{2022}

\unnumbered

\begin{document}

\title[\small Assessing Planetary Complexity using Epsilon Machines]{Assessing Planetary Complexity and Potential Agnostic Biosignatures using Epsilon Machines}

\author*[1,2]{\fnm{Stuart} \sur{Bartlett}}\email{sjbart@caltech.edu}
\author[1]{\fnm{Jiazheng} \sur{Li}}
\author[1,3]{\fnm{Lixiang} \sur{Gu}}
\author[4]{\fnm{Lana} \sur{Sinapayen}}
\author[1]{\fnm{Siteng} \sur{Fan}}
\author[5]{\fnm{Vijay} \sur{Natraj}}
\author[5]{\fnm{Jonathan} \sur{Jiang}}
\author[5]{\fnm{David} \sur{Crisp}}
\author[1,5]{\fnm{Yuk} \sur{Yung}}

\affil*[1]{\orgdiv{Division of Geological and Planetary Sciences}, \orgname{California Institute of Technology}, \orgaddress{\street{1200 E California Blvd}, \city{Pasadena}, \postcode{91125}, \state{California}, \country{United States}}}

\affil[2]{\orgdiv{Earth-Life Science Institute}, \orgname{Tokyo Institute of Technology}, \orgaddress{\city{Tokyo}, \country{Japan}}}

\affil[3]{\orgdiv{Department of Atmospheric and Oceanic Sciences}, \orgname{Peking University}, \orgaddress{\city{Beijing}, \postcode{100871}, \country{China}}}

\affil[4]{\orgname{Sony Computer Science Laboratories}, \orgaddress{\city{Kyoto}, \country{Japan}}}

\affil[5]{\orgdiv{Jet Propulsion Laboratory}, \orgname{California Institute of Technology}, \orgaddress{\city{Pasadena}, \postcode{91109}, \state{California}, \country{United States}}}

\abstract{We present a new approach to exoplanet characterisation using techniques from complexity science, with potential applications to biosignature detection. This agnostic method makes use of the temporal variability of light reflected or emitted from a planet. We use a technique known as epsilon machine reconstruction to compute the statistical complexity, a measure of the minimal model size for time series data. We demonstrate that statistical complexity is an effective measure of the complexity of planetary features. Increasing levels of qualitative planetary complexity correlate with increases in statistical complexity and Shannon entropy, demonstrating that our approach can identify planets with the richest dynamics. We also compare Earth time series with Jupiter data, and find that for the three wavelengths considered, Earth's average complexity and entropy rate are approximately $50\%$ and $43\%$ higher than Jupiter's, respectively. The majority of schemes for the detection of extraterrestrial life rely upon biochemical signatures and planetary context. However, it is increasingly recognised that extraterrestrial life could be very different to life on Earth. Under the hypothesis that there is a correlation between the presence of a biosphere and observable planetary complexity, our technique offers an agnostic and quantitative method for the measurement thereof.
}

\keywords{Complexity, Exoplanet, Epsilon machine, Agnostic Biosignatures}

\maketitle

\section{Introduction}
Detecting extraterrestrial life is the superlative aspiration of astrobiology. Endeavors in this field have sought unambiguous molecular or chemical signals that could positively identify a second biosphere beyond Earth. However, many such chemical fingerprints rely upon a degree of similarity between alien biochemistry and terran biochemistry, and most require additional contextual information \citep[e.g.,][]{kiang2018exoplanet,meadows2018exoplanet}. We cannot ignore the strong possibility that alien life may be chemically and materially different from life as we know it \citep{bartlett2020defining}, and this has led some researchers to consider alternative and agnostic biosignatures \citep{dorn2011monomer,guttenberg2021classification,johnson2018fingerprinting,krissansen2016detecting,krissansen2018disequilibrium,marshall2021identifying,sole2004large,walker2018exoplanet}. Here we present the first step towards a new and agnostic approach based on information theory and complexity science. The basic concept is to use the measured complexity of signals from an exoplanet as a proxy for the complexity of the physical processes occurring on that planet, potentially including the presence of a biosphere. To this end, we present a foundational study that uses data from the Deep Space Climate Observatory (DSCOVR) \citep{marshak2018earth} to assess how the complexity of Earth changes as its surface and atmospheric features are simplified and reduced. We find that the metric known as statistical complexity can indeed quantify planetary differences, and when combined with the Shannon entropy rate, a clear trend of increasing values with increasing planetary complexity is revealed.\par
The practical situation that we are targeting is the acquisition of multi-wavelength time series reflectance data from exoplanets that have no spatial element (time series from multiple wavelength channels). Our recent papers have addressed ways in which such non-spatial information could be used to infer orbital, rotational, surface, atmospheric, and even spatial details about a distant exoplanet \citep{fan2019earth,gu2020earth,jiang2018using}. Simultaneously, many authors in the agnostic biosignature realm are embracing the fact that extraterrestrial life may be compositionally different to terrestrial life. Hence there is a pressing need for biosignature frameworks that are minimally Earth-biased, which is a deeply challenging philosophical endeavor. These agnostic approaches are focusing particularly on a) the complexity of molecular components \citep{dorn2011monomer,guttenberg2021classification,marshall2021identifying} and b) the complexity of chemical reaction networks \citep{sole2004large,walker2018exoplanet}. Such frameworks are motivated by the observation that the primary distinguishing feature of life from non-life is the fact that living systems make use of information to direct the operation of complex chemical and physical processes \citep{baluvska2016having,davies2016hidden,farnsworth2013living,tkavcik2016information,walker2016informational,Witzany2020what}. Hence, the more complex and perhaps the further from equilibrium a molecule or reaction network is, the more likely it is that it derived from a biotic process.\par
In conceiving the present work, we considered whether it might be possible to use purely temporal data from an exoplanet to estimate its probability of hosting a biosphere? Given the near universally-agreed upon association between life and complexity, it is natural to consider whether the information content and complexity of exoplanet signals (time series) can be measured, and if so, do they correlate with the presence of life? We address this question by making the first complexity measurements of planetary data. We used reduced versions of the original Earth data to test whether qualitative decreases in complexity are reflected in decreased quantitative complexity measures. We also calculated complexity and entropy values for several Jupiter time series derived from Cassini data. For all three wavelengths analyzed, Earth's complexity and entropy values were higher than those for Jupiter.\par
The technique that was used for measuring time series complexity is known as epsilon machine reconstruction (EMR), which computes various properties, the most important being the statistical complexity and Shannon entropy rate. A brief introduction to this technique and its history are given in \hyperref[sec:meas_comp]{Methods: Measuring Complexity}, and an example epsilon machine is shown in \autoref{fig:eps_mach}. The approach that we took to measuring the complexity of planetary systems is described in \hyperref[sec:planet_comp]{Methods: The Complexity of Planets}. Technical details regarding the processing of planetary time series and their application within the EMR framework are given in \hyperref[sec:data_disc]{Methods: Input Data and Processing} and \autoref{fig:data_proc}.

\section{Results}
The complexity analysis proved to be an effective discriminator of different surface types. In particular, there were quantitative distinctions between cloudy vs cloudless synthetic Earths, and Earths with two or more surface types vs those with a single surface type. Detailed results and the various dependencies of complexity upon wavelength and planetary features are described in \hyperref{sec:compl_feat}. They highlight the potential power of computational mechanics and EMR to agnostically analyze signals from distant planets.\par
However, there are various sources of noise and wavelength dependencies that could complicate the interpretation of such assessments. A more comprehensive approach would be to analyse all wavelengths simultaneously, and indeed computational mechanics offers approaches for multi-variate datasets, as well as for continuously-valued data \citep{brodu2020discovering,marzen2017informational,sinapayen2017online}. Additionally, while statistical complexity quantifies the size of the minimal predictive model of a process, the Shannon entropy rate quantifies the degree of randomness in the process, or the rate at which `surprise' is generated \citep{shannon1948mathematical}.\par
It is generally agreed that the computational capacity of a physical system depends upon both of these aspects: dynamical richness, as well as the generation of surprise (randomness). Note that curves of statistical complexity vs entropy exhibit maxima at phase transitions in systems such as the Ising model and the period doubling route to chaos \citep{crutchfield2012between,feldman2008organization}. This has led to the suggestion that the computational capacity of a system is maximised at such transitions, as well as the idea that life exists at, and gravitates towards `the edge of chaos' \citep{bak2013nature,kauffman1996home,langton1990computation}. Note however, that it has since been recognised that there is no universal curve shape or scaling behavior for complexity-entropy diagrams \citep[see page 22 of][for details]{crutchfield2012between}. Although certain classes of complex systems seem to exhibit a characteristic peak in complexity at intermediate entropy values, a rich diversity of other complexity-entropy behaviors have been revealed in recent decades \citep{feldman2008organization}.\par
Turning these insights to the case of planetary data analysis, we calculated the Shannon entropy rate for all the time series presented in \hyperref{sec:compl_feat}. We used the occurrence frequencies of the discrete reflectance values as probabilities, and employed the standard expression: $H[X]=-\Sigma_{i=1}^np(x_i)\log p(x_i)$, where $H[X]$ is the Shannon entropy of the variable $X$. We then averaged across the ten wavelength bands. The results are shown in \autoref{fig:all_ent_sc}, which highlights the potential power of the complexity-based approach for exoplanet signal analysis.\par
The wavelength-averaged results show that on the complexity-entropy plane, despite intrinsic noise and data sampling issues, there are groupings along a monotonic slope of increasing complexity and entropy. In the lower left corner of the plane are the simplest synthetic Earths (cloudless, mono-surface worlds), and on the opposite corner the most qualitatively and quantitatively complex (the original data for the unaltered Earth). Within the cloudy worlds, there is a relatively small distinction between the mono-surface and multi-surface groups, but it is encouraging to see such a distinction in the presence of the obscuring effect of clouds.\par

\section{The Case of Jupiter vs. Earth}
\label{sec:jupiter_earth}
In addition to assessing the relationship between planetary features and time series complexity, a key goal of this study was to explore the methodology's potential for distinguishing living from non-living planets. A primary candidate for such a comparison is the gas giant Jupiter, generally assumed to be sterile and known for its complex atmosphere that exhibits both periodic and stochastic features (though the atmosphere has a relatively low level of chemical disequilibrium).\par
Jupiter has long been a focus of astronomers and has been imaged by many instruments. For the following comparison, we used data from the Cassini mission, which took images of Jupiter over a period of approximately $44$ Earth days ($\sim107$ Jupiter days). The reflectance images were resolved in nine wavelength channels, three of which were sufficiently close to channels in the Earth data for a comparison: 1) 443nm (Earth) and 450.9nm (Jupiter), 2) 551nm (Earth) and 568.2nm (Jupiter), and 3) 764nm (Earth) and 750.5nm (Jupiter). The shortest wavelength time series for Earth and Jupiter are shown in \autoref{fig:E_J_timeseries}.\par
Following processing (described in \hyperref[sec:jupiter_data]{Methods: Jupiter Data Processing}), we performed EMR on the time series corresponding to the wavelengths given above, for a range of kernel width parameter values $k$ (see \hyperref{sec:param_sens} for details on the kernel width parameter, and \autoref{fig:E_J_wl_k} for the complete set of results). \autoref{fig:E_J_ent_c} shows the Earth and Jupiter complexities when averaged over the full range of $k$ values, plotted as a function of the Shannon entropy rates of the respective time series. For each wavelength considered, Earth exhibits a higher complexity and entropy, by $\sim1.25$bits for complexity, and $\sim1$bit for entropy (though this varies with wavelength). It is possible that the lower image resolution of the Cassini data impacts the information content of the Jupiter data, therefore future work will attempt to perform this analysis using higher resolution images from the Planetary Virtual Observatory and Laboratory: \url{http://pvol2.ehu.eus/pvol2/}\par
Despite the challenges of data acquisition and processing, the above results suggest that Earth's planetary complexity is higher than that of Jupiter. Deducing the causal influences that produce this difference (atmospheric properties, the presence of a biosphere, etc.) is a significantly deeper question that is discussed below, and will be addressed by future studies. However, the present results are at least consistent with the hypothesis that planetary complexity correlates with the presence of life.

\section{Discussion}
\label{sec:disc}
% Intro, agnostic nature of method
Future missions will continue to survey a diverse range of exoplanets, and time-resolved data will be an important component of these observations. While spectral and chemical biosignature approaches continue to develop, our proposed complementary approach makes no assumptions about the molecular details of putative alien biospheres, and instead prioritises information content and complexity. These properties are potentially general and abstract, and could mitigate the risk of missing alien life hiding in plain sight \citep{bartlett2020defining}.\par
% Filtering step in exoplanet analysis tool
Our methodology could also serve as a false positive or negative filtering stage in a larger exoplanet assessment scheme. Such a scheme could rank planets in terms of whether they warrant further investigation, most likely using Bayesian inference \citep[see][for recent Bayesian approaches to biosignature estimation]{kiang2018exoplanet,walker2018exoplanet,catling2018exoplanet}. For example, imagine that oxygen is detected in a planetary atmosphere and the discovery claimed to be indicative of a biosphere. If the planet exhibits a low complexity and low entropy, we might question whether the oxygen is from a biotic source, or a byproduct of a geochemical process, since life is not just distinguished by what it produces or is made from, but by what it \textit{does}. Conversely, an observed exoplanet might exhibit no spectral or chemical features associated with life as we know it, but may show a high complexity. In this case, the planet might be worthy of further study because the high complexity would suggest a high computational capacity and physical richness. Computation and information processing are characteristic hallmarks of life that are likely to be universal, even if the molecular constituents of life vary \citep{bartlett2020defining,baluvska2016having,davies2016hidden,farnsworth2013living,tkavcik2016information,walker2016informational,Witzany2020what}.\par
% Signal to noise ratio test
In order to explore potential false positives, we performed an additional test by adding uniformly distributed noise to the data. The results, presented in \hyperref{sec:sig_noise}, showed a decreasing complexity trend as a function of noise level. As the noise infiltration tended to 100\%, the complexity and entropy values almost converged, implying that the epsilon machines were tending towards compact random number generators with similar distributions to the input data. The complexity signal degraded slowly up to a noise level of $\sim10\%$, and thereafter decreased more rapidly with noise level. Thus, we can estimate that a signal-to-noise ratio of $\sim10$ might be sufficient to retain the relevant signals that produce high complexities.\par
% False negatives
Let us now consider false negatives; might we observe a planet with low complexity that in fact fosters a biosphere? One such candidate that has been widely studied is a tidally-locked exoplanet, whose rotation period and orbital period are synchronised. Modeling studies suggest that these planets might have clouds that are locked to the sub-solar region and sub-solar to anti-solar circulations \citep{pierrehumbert2019atmospheric}. Their light curves might therefore exhibit low complexity.\par
% Simple biospheres as false negatives
One can also imagine simpler biospheres than ours, perhaps comprised only of simple, single-celled organisms, perhaps evolving or metabolising very slowly, deep below the surface. While this is a compelling possibility, the chances of detecting such a biosphere from remote measurements of \textit{any} kind seem slim. If a biosphere interacts only weakly with its environment, and if the fluxes of matter and energy through it are relatively low, its presence would probably remain hidden from chemical or spectral analyses. From a great distance we would observe what appears to be a sterile world.\par
\begin{figure}[h]
\centering
\includegraphics[width=0.9\textwidth]{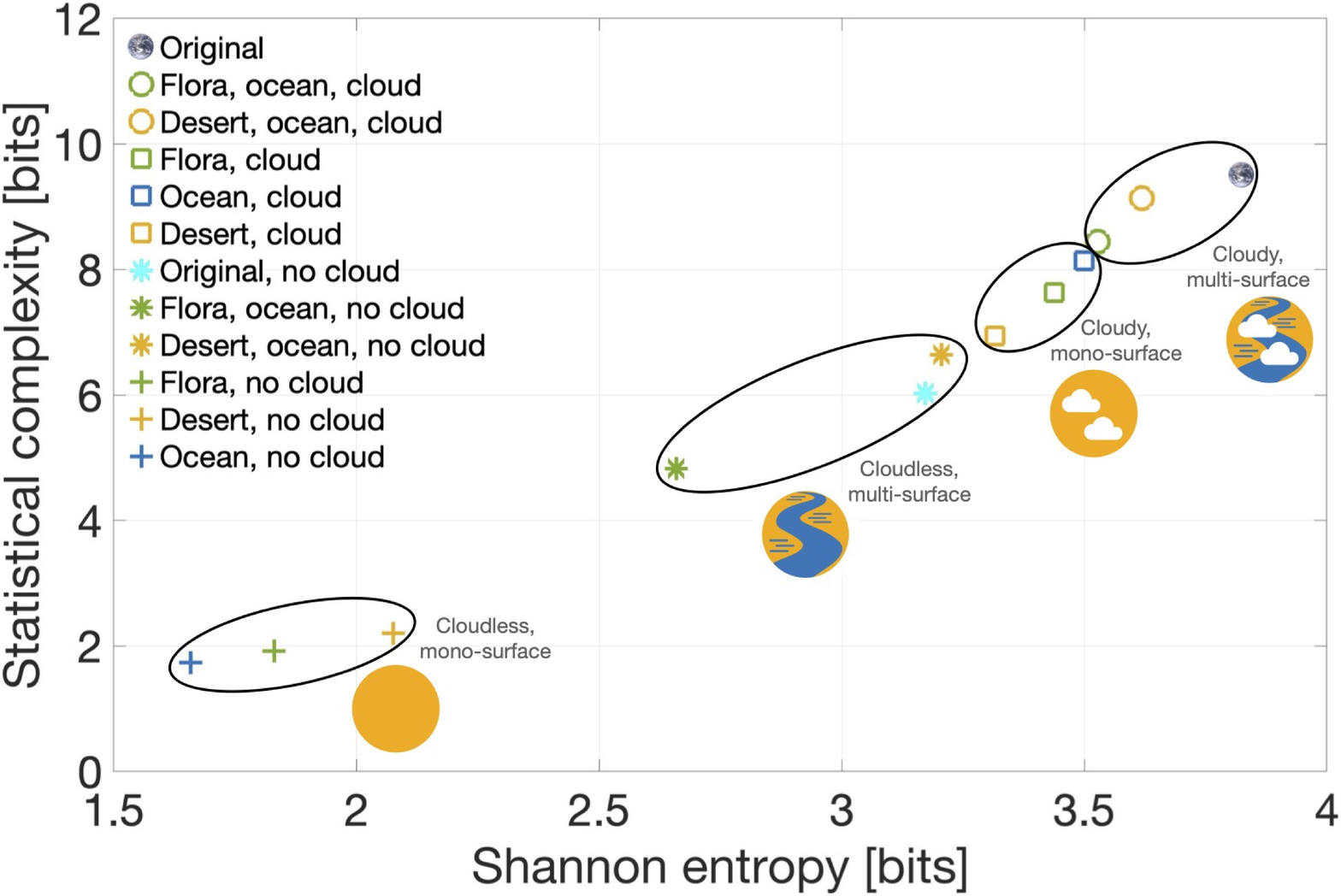}
\caption{Statistical complexity as a function of Shannon entropy averaged across all wavelength bands for the original and synthetic Earth data.}
\label{fig:all_ent_sc}
\end{figure}
\begin{figure}[h]
\centering
\includegraphics[width=0.8\textwidth]{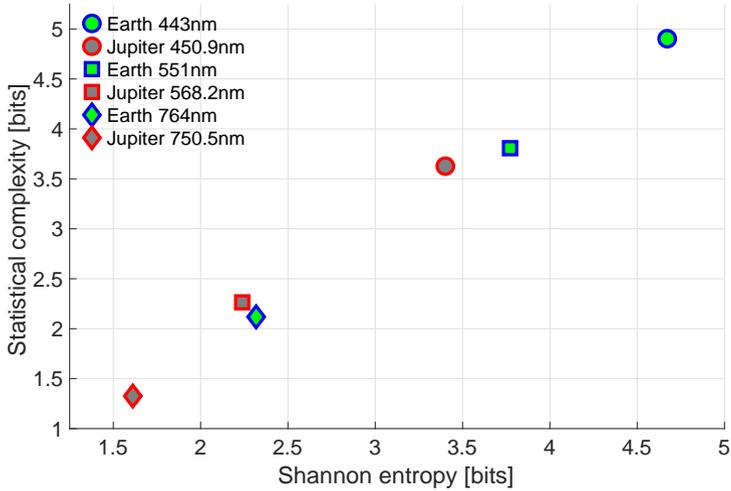}
\caption{Average complexity of Earth and Jupiter as a function of Shannon entropy for the three wavelength bands analysed.}
\label{fig:E_J_ent_c}
\end{figure}

% Other regions of the complexity-entropy plane
Our analysis showed a monotonic increase of complexity and entropy with more complex synthetic Earths. However, given that EMR can detect simple random processes, we can consider whether other regions of the complexity-entropy plane of \autoref{fig:all_ent_sc} could be occupied by different classes of planets (as suggested by the comparisons in \hyperref[sec:jupiter_earth]{The Case of Jupiter vs. Earth}, given that those results trace out a curve of slightly lower complexities than those in \autoref{fig:all_ent_sc}). New endeavors currently underway by our group are using exoplanet General Circulation Models to generate a broad range of exoplanet data. This will allow us to explore other regimes of the complexity-entropy plane, where the statistical complexity might be high, but the entropy low, for example. This might be the case for planets with complex but predictable surface features (e.g., a cloudless world with a very complicated surface). Conversely, there could be chaotic, cloud-covered worlds that exhibit almost no detectable and reproducible structure, but only stochastic features.\par
% Correlation vs. causation, life as a planetary process or parallel occurrence
If one were to observe Earth from a great distance, it would be extremely difficult to directly image explicit signatures of the biosphere. In the present work, the features contributing to Earth's measured complexity are arguably abiotic (e.g., clouds). Hence, it could be suggested that we have in fact discriminated a fast-rotating planet, with a quasi-axisymmetric (zonal) general circulation (Jupiter) and cloud fields, from a more slowly rotating planet (Earth) with a less axisymmetric circulation (high index flow) and corresponding cloud fields. Bearing this in mind, the key question becomes the degree to which the correlation between complexity and the presence of a biosphere represents a causal link (the extent to which the Earth's biosphere \textit{causes} the measurable complexity in its remotely-observable abiotic features). Although it might appear that Earth's most complex features are abiotic, they are all strongly and inextricably linked to the biosphere, with which they have been co-evolving for billions of years. We contend that life is in fact a `planetary phenomenon' \citep[see detailed discussion of this concept in][]{smith2016origin}. It seems highly unlikely that the removal or absence of life would leave the main abiotic features of Earth trivially unchanged. It seems more likely that the climate system, oceans and lithosphere would adopt very different compositions, dynamical regimes and appearance, given that the planet is not just the linear sum of its components.\par
% Life as a planetary phenomenon/Gaian bottleneck
The idea of life as a planetary phenomenon (something that happens `to' a planet rather than `on' a planet) is underscored by contemporary modelling studies of planet-biosphere coevolution and the Gaian bottleneck concept \citep{chopra2016case,lenardic2021habitability,lenton2018selection,nicholson2018gaian}. Such studies do not suggest that the Earth can be considered as a giant, single organism, but rather that selective forces at ecosystem and larger scales can favor collectives of organisms that enhance rather than degrade favorable environmental conditions. They also suggest that sustaining a biosphere in the long term may depend upon passing through a Gaian Bottleneck: a threshold level of environmental regulation by a biosphere that must be attained in order to prevent the planet from falling into an attractor state that is inhospitable to life.\par
% Life's impacts on its planetary environment
Hence, it is possible that out of the space of viable states for a planet, those that are favorable to life occupy a relatively small volume, and there is no \textit{a priori} reason to think that such small volumes are intrinsically dynamically attractive. Therefore, active regulation on the part of a biosphere might be necessary to keep a planet within this special sub-volume of its phase space. The mechanistic details of such regulation would likely include alterations of atmospheric composition \citep{kasting2002life}, which impacts weathering \citep{dyke2011towards}, atmospheric pressure \citep{li2009atmospheric}, and surface albedo \citep{harvey2015circular,wood2008daisyworld}.\par
% Sterile Earth comparison and correlation vs. causation
We cannot of course rule out the possibility that a sterile world or an equivalent non-living Earth would produce similar or higher complexity values than the actual Earth, and indeed such a result would imply that the utility of our approach lies in quantitatively assessing planetary complexity (not including the detection of biosignatures). However, until we can faithfully simulate a sterile Earth, the question of the separability of the biosphere and Earth's abiotic features remains open. In the present work we have presented a \textit{correlation} between planetary complexity and life, and look forward to future studies that can deduce the degree of \textit{causal} power that flows from the biosphere to Earth's remotely-observable abiotic features, or whether non-living planets can exhibit similar or higher complexity.\par
% Observation time scales required for our analysis
A natural question that arises from our study is the required observation time to make reasonable measurements of complexity. Intuitively, longer observation times are desirable, and reasonable measurements would likely require monthly or longer periods of measurement, but this of course would vary with the rotational and orbital period of the target exoplanet. However, our data length sensitivity analysis in \hyperref{sec:param_sens} shows that significant reductions in time series length do not impact the results. Hence our preliminary estimate is that observations in excess of several local months might be sufficient to estimate a planet's complexity.\par
% Even higher complexity, faster-paced worlds
It is also possible that biospheres more evolved than our own could exhibit even faster dynamics. If we consider the recent trends in information transfer, prediction, and planetary scale changes in the Anthropocene, it seems conceivable that the evolution of complexity is associated with a shortening of characteristic time scales \citep{west2017scale}. This appears to emerge from the enhanced learning rate of the biosphere that leads to a greater degree of environmental exploitation, which in turn causes faster perturbations (e.g., climate change), requiring even faster learning. These are fascinating open questions that we can only speculate upon until more instances of life are discovered.\par
% Comparison with Zip complexity
The nature and measurement of complexity remain open philosophical issues and EMR is one approach among many. Therefore, we compared our results to another ranking based on Zip compressibility. Overall, Zip compressibility provided weaker distinctions between the synthetic Earth types than EMR (see \hyperref{sec:zip} for details). Echoing the epsilon machine results, Earth exhibited a higher Zip complexity than Jupiter for all three wavelengths considered.\par
% Alternative metrics
Beyond EMR and Zip compressibility, there are a range of other metrics that could provide similar or superior distinguishing power, including set complexity \citep{galas2014describing}, multi-variate (multi-wavelength in this context) statistical complexity \citep{marzen2017informational,sinapayen2017online,marzen2017structure}, the complexity of the principal components of the data, or automated reconstructions of differential equation sets \citep{brodu2020discovering}. The present work represents a foundational starting point, an example of how statistical complexity can be used to distinguish levels of qualitative planetary complexity, along with a first test case in which Earth's complexity is greater than that of a sterile world. The basic idea of this paper can be extended, developed, tested, scrutinised and discussed, and we hope that from this discourse will emerge evolved and more powerful approaches.

\section{Conclusions}
We have presented an agnostic approach for exoplanet characterisation, based on methods from information theory and complexity science. The epsilon machine framework has brought insights into many enigmatic systems \citep[e.g.,][]{bertello2008application,marzen2018intrinsic,munoz2020general,park2007complexity,varn2015chaotic}, and we believe that it can also be wielded to great effect in planetary science.\par
% Earth's complexity appears to stem from a combination of a visible surface with different surface types, with a fixed period and transient clouds. The clouds in particular appear to contribute a significant fraction of the Earth's remotely-observable complexity. This aligns with intuition because clouds entail a range of both reproducible and stochastic effects. They can show significant structure and predictability, but also exhibit chaotic fluctuations that defy forecasting. In the presence of clouds, changes in land surface type only have an impact on complexity at higher wavelengths, although average complexity is distinctly lower for the mono-surface, cloudy worlds. For cloudless versions of Earth, the average complexity of multi-surface worlds is higher than those of mono-surface worlds. The results indicate the importance of observations at multiple wavelengths.\par
When combined with calculations of Shannon entropy rate, there is a clear gradient in the linear sum of complexity and entropy, with increasing complexity of planetary features. This ranking based upon the combination of complexity and entropy is in line with expectations based on a qualitative sense of complexity. The different synthetic Earth features can be grouped from simplest to most complex into the following classes: a) cloudless, mono-surface, b) cloudless, multi-surface, c) cloudy, mono-surface and d) cloudy, multi-surface. Notwithstanding intrinsic noise and sampling issues, these groups form non-overlapping regions, and are aligned in order of increasing complexity (both qualitative and statistical). We also performed a first comparison of Earth's complexity with that of a non-living world: the gas giant Jupiter. We found that peak and average values of complexity and entropy were higher for Earth, consistent with the hypothesis that planetary complexity correlates with the presence of life. This does not of course prove the hypothesis to be true; it is still possible that data from other sterile worlds may exhibit similar or higher complexities to Earth. As our research explores ensembles of data from exoplanet simulations, we will be able to continue to test the hypothesis, and further explore the various regions of the planetary complexity-entropy plane.\par
Although this work highlighted the significant role of clouds, our Jupiter-Earth comparison also raises a deeper question: what features of Earth produce its higher complexity to Jupiter? This is especially pertinent given that both planets have complex cloud systems. Clearly, having ocean and a range of land surfaces contributes to Earth's complexity, and the diversity of land spatial types is partly due to the presence of life (e.g., the various types of vegetation surface with their characteristic spectra). It is also reasonable to suggest that the cloud dynamics of Earth have been influenced by the biotic production of various chemical species such as oxygen and methane, not to mention the significant role that life plays in planetary-scale biogeochemical cycles. Deducing the precise causal links between biological processes and remotely-observable features such as clouds is a challenging endeavor, however we are confident that modern approaches, such as causal discovery and Bayesian networks, may shed light on this intriguing question.\par

%%%%%%%%%%%%%%%%%%%

% \renewcommand*{\figureautorefname}{Extended Data Fig.}
\section{Methods}
\subsection{Measuring Complexity}
\label{sec:meas_comp}
The idea of complexity has traditionally been difficult to formalise, but there is a general consensus that the complexity of the living world has increased over time, sometimes in dramatic `major transitions' \citep{adami2000evolution,lineweaver2013complexity,smith1997major}. Under the assumption that evolving life becomes more complex, the measurement of complexity becomes a key challenge. In the context of the present work, this problem is twofold: 1) How is complexity formally measured, and 2) How can we measure the complexity of an exoplanet using remotely-detectable signals?\par
\begin{figure}[h]
\centering
\includegraphics[width=0.8\textwidth]{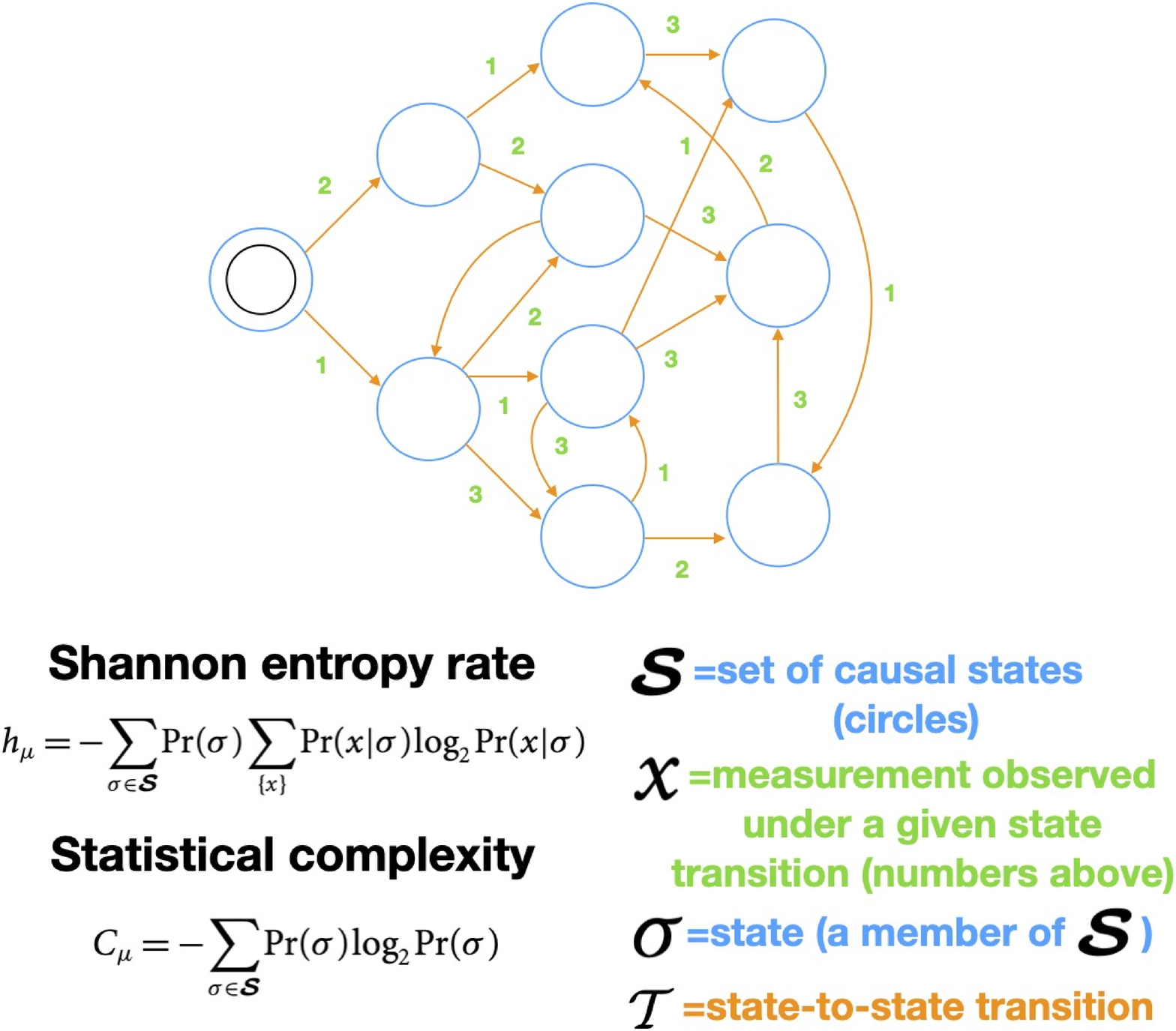}
\caption{Epsilon machine example. A simple epsilon machine with a set of discrete states (blue circles), a start state (double circle on the left), a set of state-to-state transitions (orange arrows), and a set of measurement observables for each transition (green numbers). Each transition has a certain probability of occurrence (not shown in the diagram). The Shannon entropy measures the intrinsic randomness of the model (technically the state probability-weighted information content of the transition probability distributions). The statistical complexity measures the information content of the state space of the model. Notation follows that of \citep{crutchfield2012between}.}
\label{fig:eps_mach}
\end{figure}

There have been numerous attempts to quantify complexity \citep[e.g.,][]{adami2002complexity,gell1996information,wolpert2007using}, and the most well-known is the metric due to Kolmogorov \citep{chaitin1990information,kolmogorov1963tables}. However, this measure essentially quantifies the degree of randomness in a string, and there is no universal algorithm for its computation. In contrast, an epsilon machine, described below, can represent simple random processes compactly, i.e., a completely random process can be represented by an epsilon machine with low complexity \citep[see discussion in][ for further details]{crutchfield2012between}. Simple deterministic processes can also be represented compactly (with low complexity). The advances that founded this approach began in the 1980s \citep{packard1980geometry,shalizi2001computational}. These endeavors were inspired by the basic question of how to most rationally construct models of physical systems based on limited access to their internal degrees of freedom. It was found that even simple projections of multi-variate measurement time series can reveal information about the phase space structure of non-linear systems \citep{packard1980geometry,crutchfield1986chaos}. Thanks to four decades of development, the measurement of complexity for discrete time series is now possible through rigorous, formal algorithms \citep[see][and references therein]{crutchfield2012between}. These algorithms take as input the time series being analyzed, and as output produce a minimal, optimised model, capable of reproducing time series that are statistically equivalent to the input time series. The size of such a model (specifically, the information content of its state space distribution) produces a metric known as statistical complexity (see \autoref{fig:eps_mach}).\par
The models are known as epsilon machines, and they are designed to fulfill several criteria: (i) the ability to reproduce the deterministic (reproducible) features of the given time series, (ii) the ability to reproduce the stochastic (non-reproducible) features of the given time series, (iii) the property of being optimal predictors of the processes that produced the time series, and (iv) the property of being the most compact or minimal machines for the task \citep{crutchfield2012between}. The formal process described above is known as EMR, and the first algorithm for its implementation was the causal state splitting reconstruction process \citep{shalizi2001computational}, and others have been created since \citep[e.g.,][]{brodu2011reconstruction}.

\subsection{The Complexity of Planets}
\label{sec:planet_comp}
Given that computational mechanics provides an answer to the first question raised in the previous section (the complexity of time series can be quantified by the statistical complexity), we can now turn to the second question: How can the complexity of an exoplanet be measured using remotely-detectable signals? By extension from the association between life and complexity, we can propose the following hypothesis: those planets exhibiting greatest complexity will have the highest likelihood of fostering their own biospheres. Our single example of a living planet will serve as the initial benchmark (every astrobiologist wishes the set of living planets had more than one member). Negation of this hypothesis would require the following characteristics to both be met by a planet: 1) complete sterility (lack of a biosphere), and 2) a measured complexity that was equal to or greater than that of Earth. However, even if such a negation came to pass, our suggested approach would still retain considerable utility for exoplanet analysis (see \hyperref[sec:disc]{Discussion}).\par
Intuitively, we would expect a broad spectrum of complexity levels for different planetary bodies. For example, the moon has very low complexity, since it exhibits little change over time, few periodic or stochastic features that can be observed (in time series measurements with no spatially-resolved information), and almost certainly no life. However, the Sun and Jupiter, if viewed simply by their outermost discernible features, are relatively complex. There are coherent dissipative structures in both cases, periodic, repeating features, as well as fluctuations and stochasticity. Finally, consider the only known living planet, our home. Observing Earth from a distance, we see its atmosphere, which exhibits many complex features over a broad range of timescales. These features are intrinsically linked to the atmosphere, hydrosphere, and of course the biosphere. Earth appears to be the most complex body in the solar system, and its biosphere is a primary driver of this complexity \citep{dyke2011towards,kasting2002life,lane2002oxygen,olejarz2021great}. However, until now there has been no attempt to quantify this complexity so that the aforementioned hypothesis (that Earth is the most complex body in the solar system) can be tested.\par
Thus, if we could measure the complexity of solar system bodies via remote measurements (assuming no spatial resolution, since this would be the case for distant exoplanet measurements), would the Earth exhibit the highest complexity? Development of the methodology presented here was motivated by this overarching question. Using EMR as described above, we can take time series observations from Earth and other planets, and objectively characterise them in a way that makes no prior assumptions about composition or dynamics. The resulting complexity measurements could then be used to assess planetary complexity and potentially as a proxy for the likelihood of a biosphere. We assume throughout that we only have access to non-spatial information, but that we may have time series from multiple wavelength channels.

\subsection{Input Data and Processing}
\label{sec:data_disc}
Our previous work demonstrated that various planetary properties can be extracted from multi-wavelength time series of reflected light from a planet \citep{fan2019earth,jiang2018using}. These studies used Earth as a proxy exoplanet, by taking images from the DSCOVR satellite and integrating over the planet's disk to yield a single pixel per image. This produced 10 time series, one for each of the reflected light wavelength channels. Analysis of these 10 time series (henceforth called the original time series) enabled discernment of several planetary properties including orbital period, rotation period, land fraction and ocean fraction. Surface information was found to be in the second principal component of the time series variation, while clouds contributed the maximal variation (this result was recapitulated in the complexity analysis: cloudy worlds were found to have the highest complexities).\par
In this work, we make use of Earth images that are deconstructed and reconstructed with respect to spatial types \citep[see][for details of the process]{gu2020earth}. The spatial types are categorised as high cloud, low cloud, vegetation, desert, and ocean. Representative spectra for each of these spatial types were used to reconstruct Earth images in different combinations, allowing artificial reductions in the complexity of Earth's surface. Each set of reconstructed Earth images was disk-integrated to transform them to single pixels resolved in time (time series analogous to those used in \citep{fan2019earth}). In this manner, we were able to compare synthetic versions of Earth with different combinations of spatial types.\par
\begin{figure}[h]
\centering
% \subfigure[\label{fig:orig_orig}Original]{\includegraphics[width=0.45\textwidth]{ext_data_fig2a_orig_orig.eps}}
% \hspace{5mm}
% \subfigure[\label{fig:orig_disc}Original after normalisation and discretisation]{\includegraphics[width=0.45\textwidth]{ext_data_fig2b_orig_disc.eps}}
% \subfigure[\label{fig:des_orig}Cloudless desert-world]{\includegraphics[width=0.45\textwidth]{ext_data_fig2c_des_orig.eps}}
% \hspace{5mm}
% \subfigure[\label{fig:des_disc}Cloudless desert-world after normalisation and discretisation]{\includegraphics[width=0.45\textwidth]{ext_data_fig2d_des_disc.eps}}
% \subfigure{\includegraphics[width=0.95\textwidth]{ext_data_fig2_legend.jpeg}}
\subfigure{\includegraphics[width=0.95\textwidth]{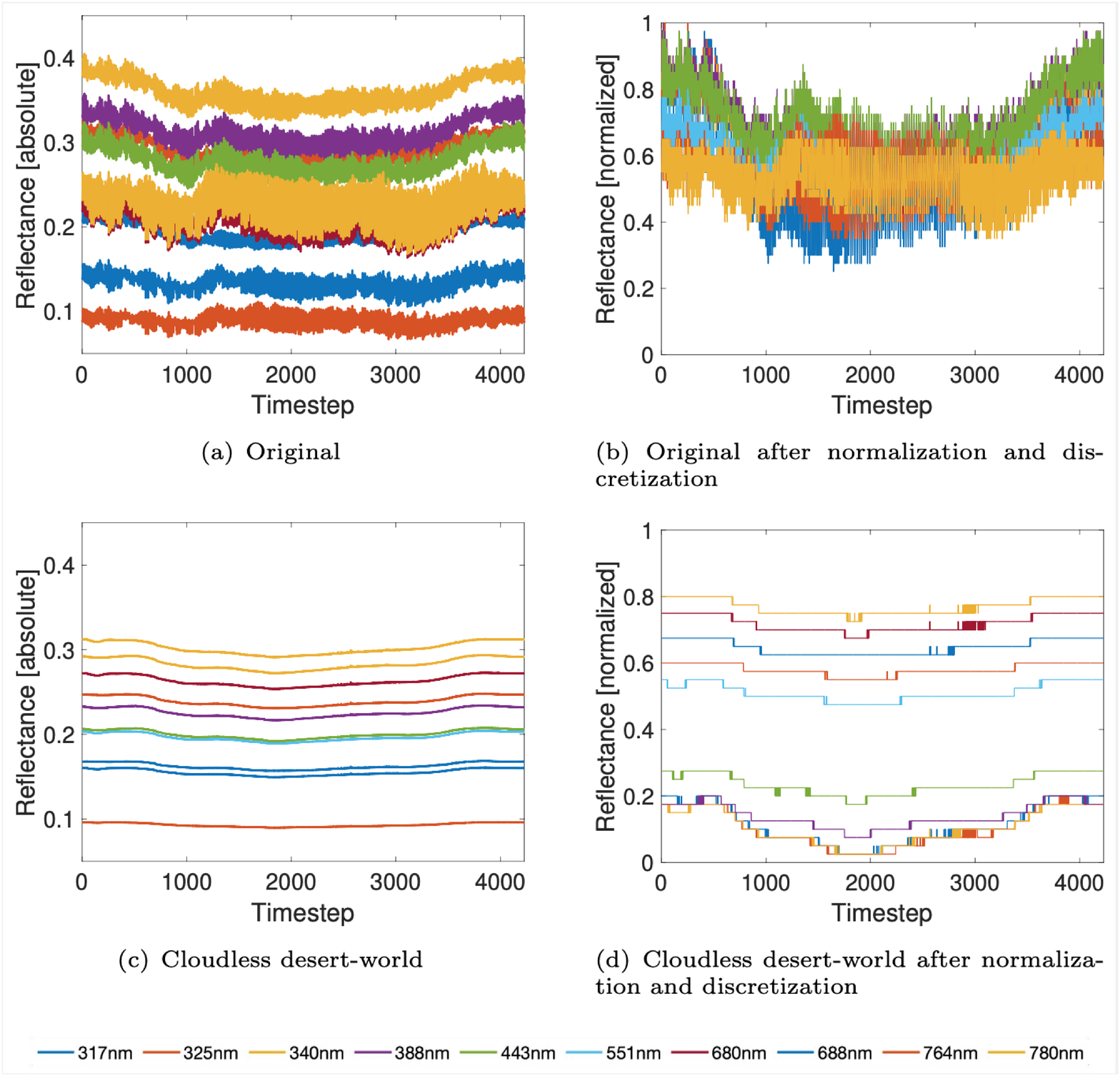}}
\caption{Time series processing. Examples showing the pre-processing of time series in preparation for EMR. a) Original, unaltered time series from the 10 reflectance channels, b) Original time series after normalisation and discretisation, c) Synthetic time series generated by replacing all pixels with a characteristic desert spectrum, d) Normalised and discretised versions of the series in (c).}
\label{fig:data_proc}
\end{figure}

\autoref{fig:data_proc} shows the original time series from the DSCOVR data in panel (a), and synthetic time series computed by replacing all image pixels with characteristic spectra for desert, and disk-integrating to one pixel in panel (b). As expected, the cloudless desert-world exhibits very little temporal variation. The reason there is any variation at all (given the homogeneous surface in this case) is the Sun-Earth distance change, and because the spectra have a viewing geometry dependency.\par
In order to assess the complexity of such time series, they must be discretised, since epsilon machines consist of a finite number of discrete states and transitions \citep[see][for further details on data requirements for EMR]{crutchfield2012between,shalizi2001computational,brodu2011reconstruction}. We adopted a simple normalisation procedure that uses the maximum and minimum reflectances from each wavelength band across all spatial type reconstructions:
$r^n_{\lambda,s,t}=\frac{r_{\lambda,s,t}-r_{\lambda_{min}}}{r_{\lambda_{max}}-r_{\lambda_{min}}}$, where $r^n_{\lambda,s,t}$ is the normalised reflectance for wavelength $\lambda$, spatial type reconstruction $s$ and time $t$, $r_{\lambda,s,t}$ is the original value, and $r_{\lambda_{min}}$ and $r_{\lambda_{min}}$ are the minimum and maximum reflectance values for wavelength $\lambda$ across all the different spatial type reconstructions. Note that normalisation is not technically necessary the EMR algorithm is indifferent to the numerical values of the input series). However, we normalise in this work to ensure consistent comparisons and to aid visualisation.\par
It is essential to use data with a finite set of values for EMR, hence after normalisation, the data values are discretised to a resolution of 40 levels (these are relative reflectance levels). This resolution choice of 40 levels is a simple trade-off between representation of temporal features and computation time required for EMR, hence higher resolutions could also be used (see \hyperref{sec:param_sens} for a discretisation sensitivity analysis).\par
\autoref{fig:data_proc} illustrates the normalisation and discretisation process for two spatial compositions: the original dataset (top row), and a cloudless desert-world version of Earth (bottom row), one of the least complex time series assessed. The discrete time series can then be used as input for EMR. The `Timeseries' algorithm of \citep{brodu2011reconstruction} was used for this task, with a kernel width parameter of $k=2.5\times10^{-4}$. This value provided the most effective numerical discrimination in terms of complexity values, among those tested (see \hyperref{sec:param_sens} for a $k$ value sensitivity analysis).\par
Note that there are missing points in the DSCOVR dataset, leading to uneven time spacing (the average sampling rate is bi-hourly and the dataset covers a time period of one year). This will be addressed with appropriate missing data methods in future analyses, and we anticipate that such upgrades will lead to concomitant improvements in the results. However, for the present work, the only extra pre-treatment conducted on the data was the replacement of three erroneous data points with their linearly interpolated counterparts. The above processing steps produced time series that could be used as input for the `Timeseries' EMR code described in \cite{brodu2011reconstruction}.\par

\subsection{Jupiter Data Processing and Analysis}
\label{sec:jupiter_data}
The Cassini images of Jupiter were disk-integrated to a single pixel as they were for Earth, producing a set of nine time series (only three were close enough to Earth time series wavelengths to be viable for comparison). These time series had a coarser resolution and shorter duration than those for Earth, making it necessary to shorten and resample the Earth series such that the two datasets had identical resolution and length. We took the respective rotation periods to be the elementary time units, and hence both sets of data were processed to have the same number of local days (107 Earth days and 107 Jupiter days) and the same number of data points (364). This was readily achieved through interpolation for the higher resolution Earth data.\par
\begin{figure}[h]
\centering
\includegraphics[width=0.8\textwidth]{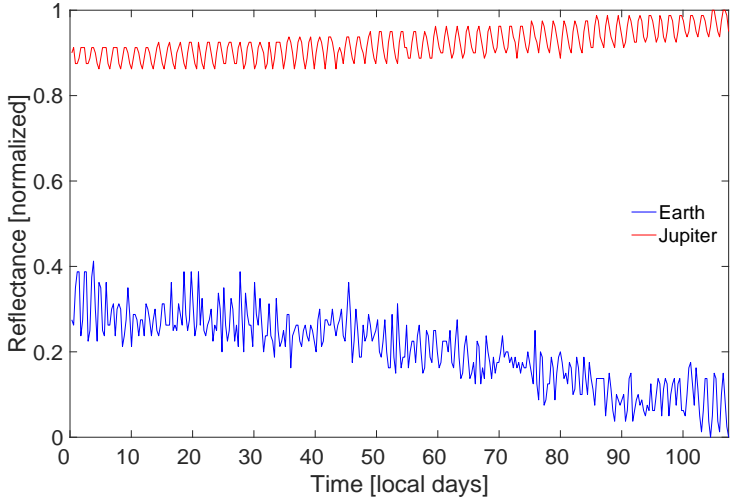}
\caption{Earth, Jupiter time series. Time series for Earth (blue, 443nm wavelength) and Jupiter (red, 450.9nm wavelength) used for the complexity comparison.}
\label{fig:E_J_timeseries}
\end{figure}

\autoref{fig:E_J_timeseries} displays the two $\sim445nm$ time series after normalisation and discretisation. For this analysis, discretisation was performed with 80 levels, due to the numerical separation between the two datasets (high reflectance values for Jupiter and low values for Earth), and relatively narrow variation of reflectance values.\par
\begin{figure}[h]
\centering
\includegraphics[width=0.95\textwidth]{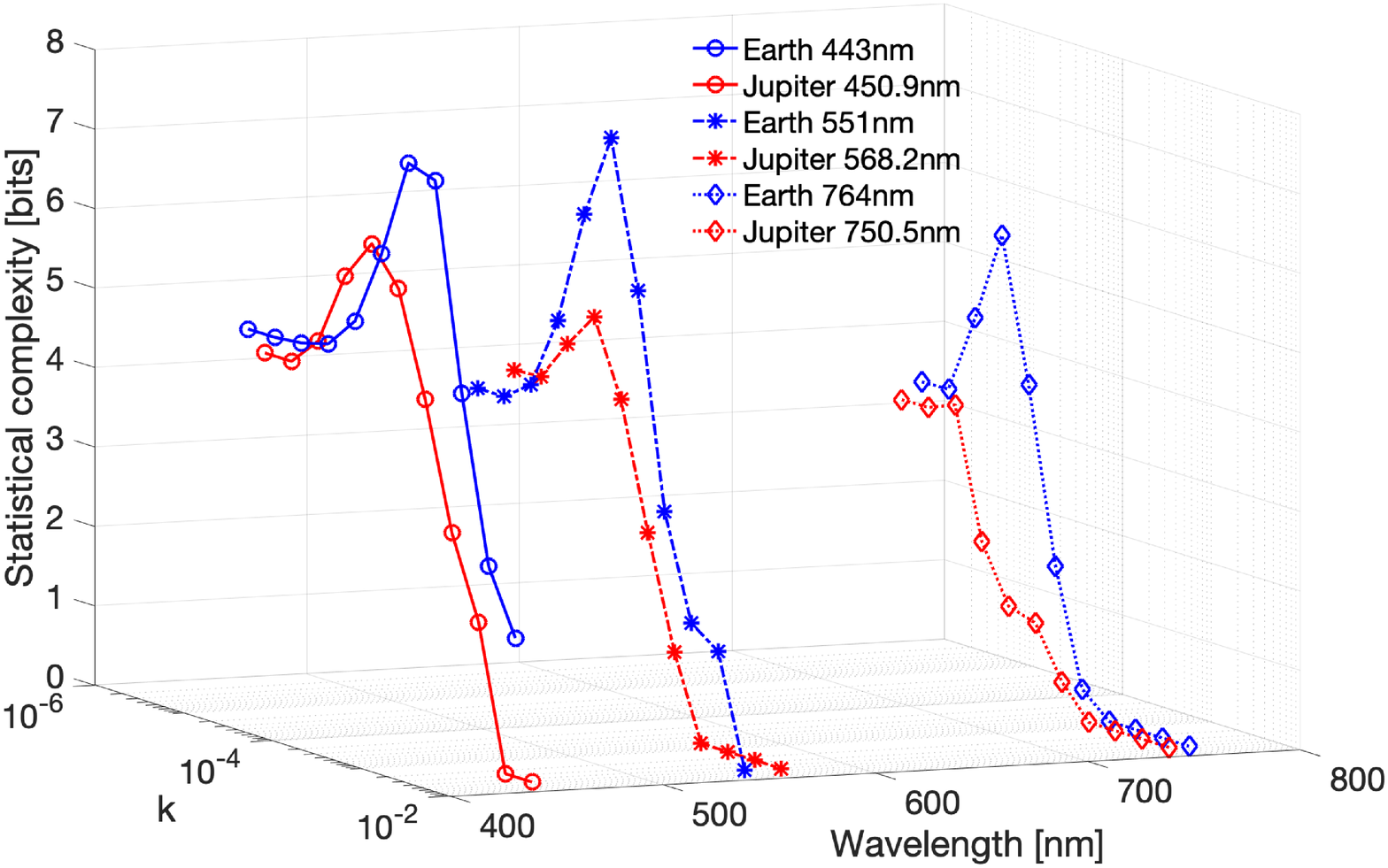}
\caption{Earth, Jupiter complexity comparison. Statistical complexities of Earth and Jupiter time series in three wavelength bands for a range of kernel width parameter values.}
\label{fig:E_J_wl_k}
\end{figure}

We performed EMR on the six (three each for Earth and Jupiter) time series described above for a range of kernel width parameter values. The results are shown in \autoref{fig:E_J_wl_k}. Note that high values of $k$ lead to a smoothing effect and epsilon machines that tend to a size of 1 state (0 complexity). Low values of $k$ lead to over-fitting, wherein the algorithm struggles to find structure in fine-scale, stochastic features, and hence produces small, stochastic epsilon machines. Intermediate $k$ values produce the most effective distinctions between the datasets, illustrated by the complexity peaks in \autoref{fig:E_J_wl_k} for both Earth and Jupiter. The crucial result of \autoref{fig:E_J_wl_k} is that for all three wavelengths considered, Earth has higher peak and average complexities.\par

%%%%%%%%%%%%%%%%%%%%%%%%%%%%%%%%%%%%%

\backmatter
\bmhead{Acknowledgments}
We acknowledge partial funding support from the NASA Exoplanet Research Program NNH18ZDA001N-2XRP. A portion of this research was carried out at the Jet Propulsion Laboratory, California Institute of Technology, under a contract with the National Aeronautics and Space Administration (80NM0018D0004). Y. L. Y. was supported in part by an NAI Virtual Planetary Laboratory grant from the University of Washington. We thank the members of the Caltech GPS `Astrobiothermoevoinfo' reading group for the various inspiring discussions that have helped catalyze ideas such as those presented here. We also thank Tricia Ewald at Caltech for valuable help with the processing of Jupiter data from Cassini. Finally, SJB would like to thank Professor Seth Bullock for being his guide into the world of complexity.

\bmhead{Author Contributions}
SJB conceived of the idea of using EMR to analyze planetary complexity and the hypothesised correlation between planetary complexity and the presence of life. He performed the complexity analysis, produced the figures and wrote the manuscript. JL provided the Jupiter Cassini data. LG and SF produced the synthetic Earth, recomposed data sets. LS assisted with the complexity analysis, results interpretation, literature review and manuscript editing. VN assisted with results interpretation and manuscript editing. JJ, DC and YY provided essential guidance, assistance with data provision, results interpretation and manuscript editing.

\bmhead{Statement of Competing Interests}
The authors declare no competing interests.

\section*{Data availability}
Source data for all time series as well as the data points in the figures in the main text are provided with this paper.

\section*{Code availability}
C++ code for the epsilon machine reconstruction process used in this study can be accessed here: \url{https://nicolas.brodu.net/recherche/decisional_states/index.html}.

% \bibliography{dscovr_compl}

%%%%%%%%%%%%%%%%%%%%%%%%

%%%%%%%%%%%%%%%%%%%%%%%%

%%%%%%%%%%%%%%%%%%%%%%

\clearpage

{\Huge{Supplementary Information}}
\section{Contributions to Complexity from Planetary Features}
\label{sec:compl_feat}
\begin{figure}[h!]
\centering
\includegraphics[width=0.7\textwidth]{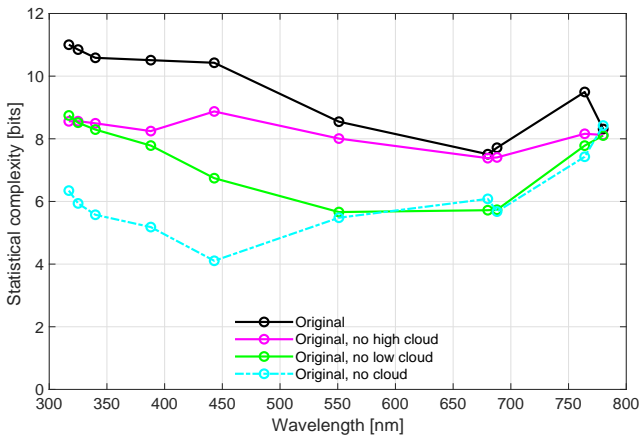}
\caption{Statistical complexity as a function of wavelength for the synthetic Earth time series with modified clouds. The black, red, blue and green curves show values for the original, no high cloud, no low cloud, and cloudless time series, respectively.}
\label{fig:cld_on_off}
\end{figure}

In this section we present the statistical complexity values computed for all synthetic Earth types as a function of wavelength. \autoref{fig:cld_on_off} illustrates the influence of clouds on statistical complexity, plotted as a function of wavelength. The original (unaltered) data exhibits the highest complexities as expected, followed by the `no high cloud', `no low cloud', and cloudless versions. Removing high clouds had little effect on complexity at longer wavelengths (overlap between cyan and green curves). There is an anomalous result at 680nm, where removal of high clouds from the `no low cloud' version actually increased the complexity. Note that at this wavelength, removing high clouds from the original time series also caused a negligible decrease in complexity. Hence at these intermediate wavelengths, the majority of the complexity decrease stemmed from the removal of low clouds.\par
The results at 780nm are ambiguous, where the complexities of all four time series are essentially equal. The distinctions are clearer at lower wavelengths such as 443nm. In the UV we see that removal of high or low clouds causes a similar reduction in complexity, and further cloud removal causes an additional decrease.\par
\begin{figure}[h!]
\begin{center}
\subfigure[\label{fig:oc_on_ff_cld}Effect of number of surface types in the presence of clouds]{\includegraphics[width=0.7\textwidth]{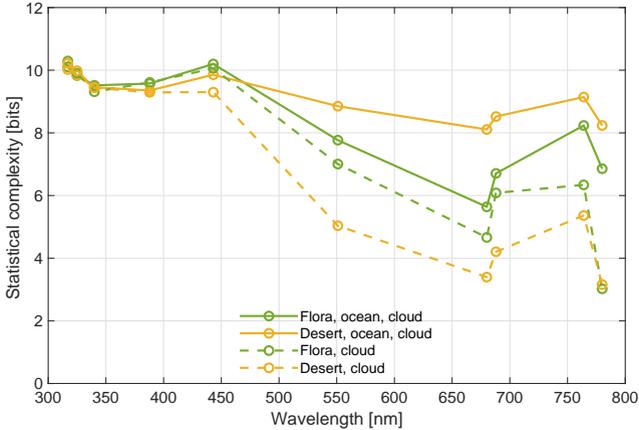}}
\subfigure[\label{fig:oc_on_ff_ncld}Effect of number of surface types in the absence of clouds]{\includegraphics[width=0.7\textwidth]{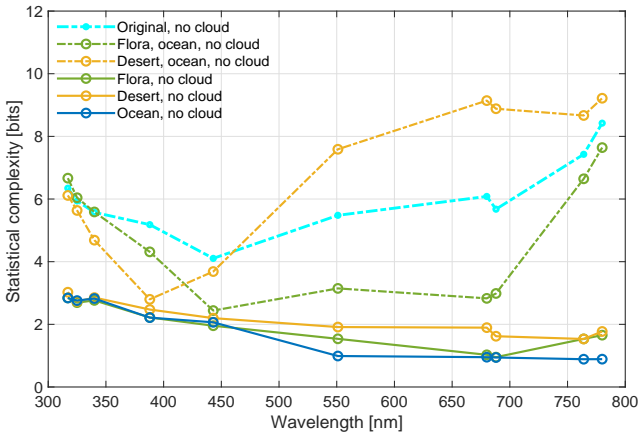}}
\caption{Statistical complexity as a function of wavelength for synthetic Earth time series. a) Surface reconstructions in which clouds are left unaltered but surface types are varied, b) Surface reconstructions in which clouds are removed and surface types are varied.}
\label{fig:comp_s}
\end{center}
\end{figure}

We now turn to the influence of the number of surface types for cloudy synthetic Earths. \autoref{fig:oc_on_ff_cld} shows statistical complexity as a function of wavelength for cloudy systems with two surface types (solid green line: vegetation and ocean surface, solid yellow line: desert and ocean surface), and one surface type (dashed green line: pure vegetation surface, dashed yellow line: pure desert surface). We see that in the UV wavelengths, the analysis makes no distinction between the two classes, which is likely due to ozone and Rayleigh scattering obscuring any information from surface features. At longer wavelengths, the number of surface features does have an impact on complexity, due to the lack of absorption in these bands (implying visibility in between clouds), though at 688nm and 764nm, absorption by atmospheric O$_2$ likely obscures surface information.\par
\autoref{fig:oc_on_ff_ncld} is similar to \autoref{fig:oc_on_ff_cld}, but for cloudless synthetic Earths. The cyan dot-dash line corresponds to the original data but with all clouds removed, the green dot-dash line corresponds to only vegetation and ocean, and the yellow dot-dash line corresponds to a desert and ocean world. On average, these cloudless, multi-surface worlds have a higher complexity than the second set, the solid green, yellow and blue lines, corresponding to cloudless flora, desert and ocean worlds, respectively. This is likely due to the large spatio-temporal contrast in signal between land and ocean for the multi-surface scenarios.\par
Note that intuitively, one might expect the vegetation-dominated (flora) worlds to exhibit higher complexities than similar abiotic versions such as a desert world. However, when the synthetic Earths are recomposed, the relevant surface image pixels are simply replaced by the respective typical spectra values (such as that for vegetation). Therefore there are no seasonal cycles or other biotic effects in the flora worlds, which are thus no more or less biological than the other synthetic datasets.\par
Overall, the removal of clouds causes a marked decrease in complexity, in line with expectation. The effects of cloud type changes were more noticeable at shorter wavelengths. It is also clear that multi-surface worlds have greater complexity than mono-surface worlds, with the discrimination being stronger at longer wavelengths. This indicates a greater contrast between surface types at longer wavelengths. The subtle wavelength-dependent effects are likely due to absorption features (or lack thereof) due to O$_2$ and ozone. Future work will explain these effects with the help of radiative transfer models.\par

\section{Parameter and Data size Sensitivity Analysis}
\label{sec:param_sens}
The kernel width parameter $k$, is an important input to the epsilon machine reconstruction code used in this work \citep{brodu2011reconstruction}. It essentially quantifies the size of the time window over which the algorithm attempts to build a maximally predictive model. Therefore, higher values tend to have a smoothing effect, and eventually, at sufficiently high values of $k$, the resulting epsilon machine reduces to a single state, single value machine. Thus, excessively high values tend to cause over-smoothing and the algorithm mistakes signal for noise. At the opposite extreme, very small values of $k$ emphasise details at finer timescales. The algorithm may thus mistake noise for data, and make spurious attempts to find patterns in that noise. At excessively low $k$ values the complexity begins to decrease, since the algorithm will revert to simple, purely stochastic epsilon machines as an attempt to match the fine-scale fluctuations.\par
\begin{figure}[h!]
\begin{center}
\subfigure[\label{fig:k_comp}Effect of kernel width parameter on statistical complexity]{\includegraphics[width=0.8\textwidth]{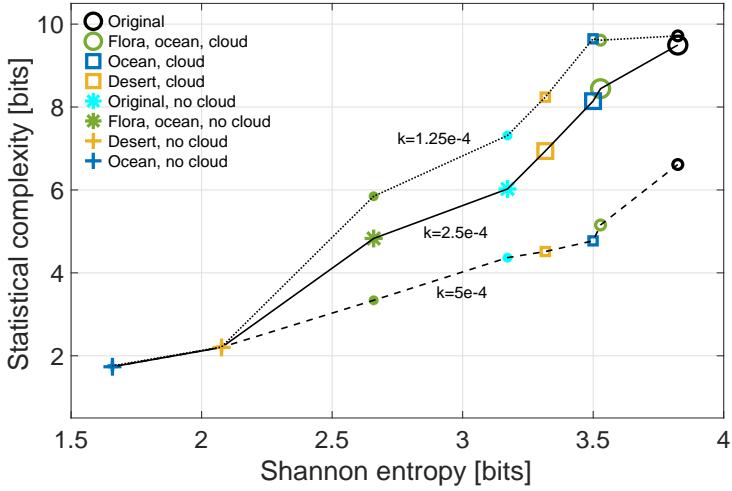}}
\subfigure[\label{fig:bins_comp}Effect of discretisation bin size on statistical complexity and entropy]{\includegraphics[width=0.8\textwidth]{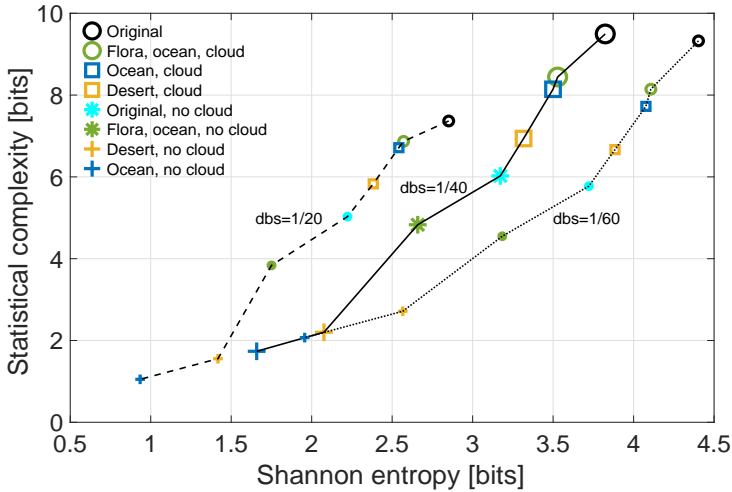}}
\caption{Statistical complexity as a function of Shannon entropy for three values of a) the kernel width parameter $k$, and b) the discretisation bin size.}
\label{fig:sens_ans}
\end{center}
\end{figure}

In \autoref{fig:k_comp} we plot the mean complexity and entropy values for a subset of our time series at three different $k$ values. The smaller $k$ value results in higher complexities in general, since the algorithm is attempting to fit finer timescale features in the data. At the higher value of $k=5e-4$, we see more of a smoothing effect and smaller complexity values. In this case the algorithm is prioritising features at coarser timescales. Despite these variations, it is clear that the general trend shown in Main Text Fig. 5 is unaffected by a doubling or halving of $k$, and our analysis found that $k=2.5e-4$ makes the most effective discriminations between the surface types. Note that the simplest synthetic Earths, the cloudless monosurface versions, are essentially unaffected by the choice of kernel width parameter. This is due to the fact that there is little to no structure in those time series for the algorithm to detect, hence changes in $k$ have little impact on complexity.\par
\begin{figure}[h!]
\centering
\includegraphics[width=0.8\textwidth]{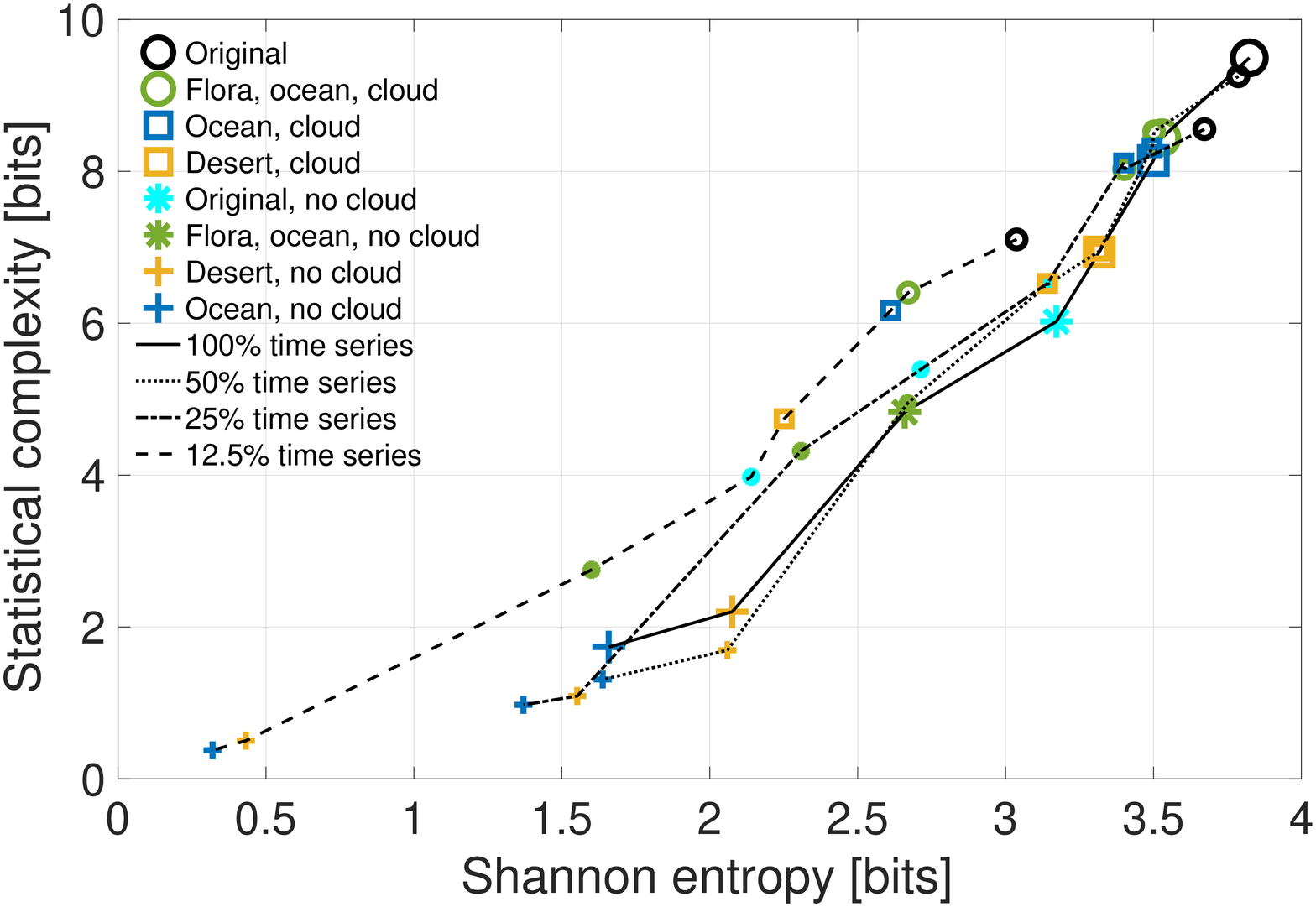}
\caption{Statistical complexity as a function of Shannon entropy for a subset of synthetic worlds with reduced data lengths.}
\label{fig:length_comp}
\end{figure}

All input time series also have to be discretised to a finite number of values. Smaller discretisation bin sizes reduce coarse-graining of fine details but incur a higher computational cost. The ideal choice of bin size is a simple trade-off between retaining sufficient detail while allowing feasible computation times. In \autoref{fig:bins_comp}, the effect of bin size is illustrated for a subset of the time series. As expected, a coarser discretisation (larger bin size) causes a reduction in both entropy and complexity due to the averaging and smoothing effect of larger bins. Using a finer resolution (smaller bin size) tends to increase the entropy but has little effect on complexity. This suggests that our chosen resolution of 40 discrete levels is sufficient to retain the primary features of the time series, since a higher resolution seems to primarily increase the level of stochasticity. As with varying $k$, it is clear that discretisation bin size does not impact the general trend illustrated in Main Text Fig. 5.
The final phase of our sensitivity analysis compared time series that had been artificially shortened by various fractions. This is a crucial consideration for future applications of our methodology, since long term observations of planetary bodies is technically challenging. We compared time series that had been reduced to a relative length of $50\%$, $25\%$, and $12.5\%$, and the results are shown in \autoref{fig:length_comp}.\par
We see that the complexity and entropy values do not show significant changes (reductions) until the data is reduced in length by at least $75\%$. This suggests that the complexity of these time series is primarily contained within stochastic, rather than reproducible, deterministic or long term features. This can be understood given that the full time series are one year in length, and hence do not contain annual cycles, but only contain diurnal and perhaps other cycles. Since there are so many stochastic features in atmospheric and planetary data such as those used here, most of the structure of the constructed epsilon machines represents an attempt by the algorithm to capture those stochastic features. The ability to find structure in stochastic data (where human eyes might just see noise) is an important ability of EMR.

\section{Signal-Noise Sensitivity Analysis}
\label{sec:sig_noise}
\begin{figure}[h!]
\centering
\includegraphics[width=0.8\textwidth]{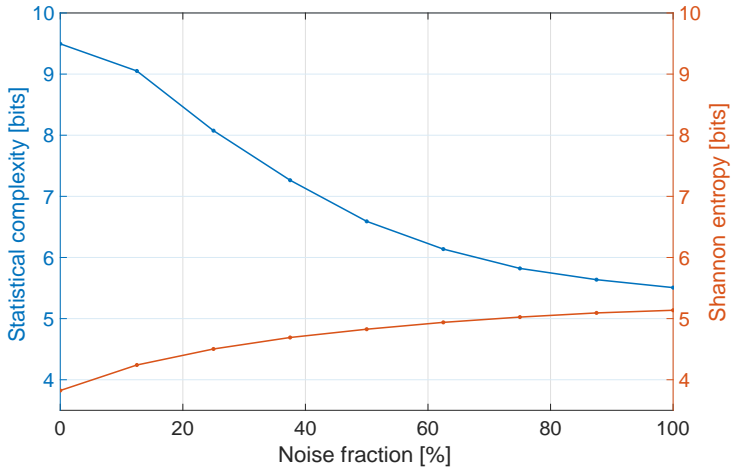}
\caption{Effect of increasing levels of artificial noise on the complexity and entropy of the `Original' Earth time series.}
\label{fig:sig_noise}
\end{figure}

Although our original data contains both intrinsic (natural) and systematic (measurement) noise, it is also important to understand how the time series complexity and entropy values change as the original signal becomes increasingly corrupted. We can readily decrease the signal-to-noise ratio by artificially replacing data points with uniformly distributed (across the unit interval $[0,1]$) random values. We introduce the continuously-valued noise after normalisation but before discretisation. We performed such a noise sensitivity study on the `Original' Earth time series and the results are shown in \autoref{fig:sig_noise}.\par
We see that that increasing noise fractions gradually degrade the signal and the complexity decreases, while the entropy increases. The EMR algorithm is unable to find structure at higher noise levels and hence the resulting epsilon machines reduce to simple random number generators with distributions that converge towards the distribution of the data (this is why the complexity and entropy converge as the noise fraction tends to 100\%). It is likely that a slight change (probably an increase) in kernel width parameter would yield exact convergence between complexity and entropy values at the 100\% noise level.

\section{An Alternative Metric based on the Zip Algorithm}
\label{sec:zip}
We have presented a complexity analysis based on Epsilon Machine Reconstruction (EMR), but this is one among many such metrics, and measuring complexity is still far from being a settled mathematical formalism. Most complexity metrics tend to be somewhat specific, designed for a certain field or set of applications. In contrast, EMR can be applied to a range of data types, even continuously-varied, multi-variate datasets \citep{brodu2020discovering,marzen2017informational,sinapayen2017online,marzen2017structure} (we are currently employing such frontier techniques to further advance the methodology presented here). One can still ask whether other measures might give similar results to EMR. As mentioned previously, Kolmogorov complexity is essentially a measure of randomness and hence compressibility, but does not have a universal method of computation. A much more common tool is the zip algorithm, used countless times every day for reducing the sizes of files without information loss. This robust algorithm can be used to measure how repetitive or compressible a volume of data is. In the realm of computer science, this notion of minimum description length is directly associated with complexity, which has led to the use of the zip algorithm as a metric thereof \citep{avinery2019universal}. In order to compare this alternative approach, we zip-compressed all the raw data files used in our analysis, and compared the resulting file sizes. The results are shown in \autoref{tab:zip_ranks}.\par
\begin{table}[h]
    \centering
    \begin{tabular}{ | l | l | l |}
    \hline
    Zip rank & Synthetic Earth type & File size [kB]\\ \hline
    1 & Original & 167 \\ \hline
    2 & Desert, ocean, cloud & 165\\ \hline
    =3 & No high cloud & 164\\ \hline
    =3 & No low cloud & 164\\ \hline
    =3 & Flora, ocean, cloud & 164\\ \hline
    =6 & Desert, ocean, no cloud & 163\\ \hline
    =6 & Ocean, cloud & 163\\ \hline
    =8 & No cloud & 162\\ \hline
    =8 & Flora, cloud & 162\\ \hline
    9 & Desert, cloud & 160\\ \hline
    10 & Flora, ocean, no cloud & 159\\ \hline
    =11 & Desert, no cloud & 136\\ \hline
    =11 & Flora, no cloud & 136\\ \hline
    12 & Ocean, no cloud & 128\\
    \hline
    \end{tabular}
    \caption{Alternative ranking based on zip compressibility. Synthetic Earth types are ranked in order of decreasing zip-compressed file size.}
    \label{tab:zip_ranks}
\end{table}
We can compare this ranking to the entropy values shown in Main Text Fig. 5, since both are measures of randomness. Although the zip algorithm ranks are approximately similar to the EMR approach, two thirds of the synthetic Earth types lie within $\sim0.6\%$ of one another, in terms of zip file size. Hence this approach does not provide a strong numerical discriminator of the surface recompositions.\par
We can also apply this approach to the Jupiter-Earth comparison. The compressed file sizes are shown in rank order in \autoref{tab:E_J_zip}. As with statistical complexity, Earth ranks higher overall.\par
\begin{table}[h]
    \centering
    \begin{tabular}{ | l | l | l |}
    \hline
    Zip rank & Time series & File size [B]\\ \hline
    1 & Earth 443nm & 569 \\ \hline
    2 & Earth 551nm & 514\\ \hline
    3 & Earth 760nm & 414\\ \hline
    4 & Jupiter 450.9nm & 408\\ \hline
    5 & Jupiter 568.2nm & 341\\ \hline
    6 & Jupiter 750.5nm & 319\\ \hline
    \end{tabular}
    \caption{Alternative Earth-Jupiter comparison based on zip compressibility.}
    \label{tab:E_J_zip}
\end{table}
In general, the power of EMR, and the reason we used it as the basis of our approach is that it correctly characterises stochastic noise as random but simple (low statistical complexity), whereas most compressibility-based approaches assign high complexities to such data.

%%%%%%%%%%%%%%%%%%%%%%%

\end{document}